\documentclass[twocolumn]{openjournal}

\usepackage[utf8]{inputenc}
\usepackage[english]{babel}
\usepackage{xcolor}
\usepackage{graphicx}
\usepackage{verbatim}
\usepackage{comment}
\usepackage{calrsfs}
\usepackage{textgreek}
\usepackage{color,colortbl}
\usepackage[normalem]{ulem}
\usepackage{soul}
\usepackage{amsmath}
\usepackage{amssymb}
\usepackage{natbib}
\usepackage{hyperref}

\DeclareMathAlphabet{\pazocal}{OMS}{zplm}{m}{n}

\def\be{\begin{equation}}
\def\ee{\end{equation}}

\def\gsim{\lower.5ex\hbox{\gtsima}} 
\def\lsim{\lower.5ex\hbox{\ltsima}} 
\def\gtsima{$\; \buildrel > \over \sim \;$} 
\def\ltsima{$\; \buildrel < \over \sim \;$} \def\gsim{\lower.5ex\hbox{\gtsima}} 
\def\lsim{\lower.5ex\hbox{\ltsima}} 
\def\simgt{\lower.5ex\hbox{\gtsima}} 
\def\simlt{\lower.5ex\hbox{\ltsima}}

\def\S*{$\Sigma_{\rm SFR}$}

\def\kms{{\rm km\,s}^{-1}\,}

\definecolor{apcolor}{HTML}{b3003b}
\definecolor{afcolor}{HTML}{800080}
\definecolor{lvcolor}{HTML}{DF7401}
\definecolor{mdcolor}{HTML}{01abdf} 
\definecolor{cbcolor}{HTML}{ff0000}
\definecolor{sccolor}{HTML}{cc5500} 
\definecolor{sgcolor}{HTML}{00cc7a}

\definecolor{linkcolor}{rgb}{0.0,0.3,0.5}
\hypersetup{
    unicode,
    colorlinks=true,
    linkcolor=linkcolor,
    citecolor=linkcolor,
    filecolor=linkcolor,
    urlcolor=linkcolor,
}
\DeclareGraphicsExtensions{.bmp,.png,.jpg,.pdf}
\graphicspath{{./}{figures/}{figs/}}

\providecommand{\kms}{\,\mathrm{km}\,\mathrm{s}^{-1}}

\providecommand{\simlt}{\lesssim}
\providecommand{\altaffiliation}[1]{\affiliation{#1}}

\makeatletter
\def\@hex@@Hex#1%
 {\if a#1A\else \if b#1B\else \if c#1C\else \if d#1D\else
  \if e#1E\else \if f#1F\else #1\fi\fi\fi\fi\fi\fi \@hex@Hex}
\makeatother
\definecolor{afcolor}{HTML}{b3443c}

\begin{document}

\title{Possible evidence for a pair-instability supernova nature\\of ultra-early JWST sources}

\author{Andrea Ferrara}
\email{andrea.ferrara@sns.it}
\affiliation{Scuola Normale Superiore, Piazza dei Cavalieri 7, 50126 Pisa, Italy}
\author{Stefano Carniani}
\affiliation{Scuola Normale Superiore, Piazza dei Cavalieri 7, 50126 Pisa, Italy}
\author{Takahiro Morishita}
\affiliation{IPAC, California Institute of Technology, Pasadena, CA 91125, USA}
\affiliation{Astronomical Institute, Tohoku University, 6-3 Aramaki, Aoba-ku, Sendai 980-8578, Japan}
\author{Massimo Stiavelli}
\affiliation{Space Telescope Science Institute (STScI), 3700 San Martin Drive, Baltimore, MD 21218, USA}

\begin{abstract}
Recent JWST observations have revealed a population of unexpectedly bright sources at ultra-high redshift ($z > 15$), challenging current models of early galaxy formation. One extreme example is \textit{Capotauro}, an F356W-dropout identified in the CEERS survey and initially interpreted as a luminous galaxy at $z\sim30$, but subsequently found to be variable over an $\sim 800$ day baseline. Motivated by this variability, we explore the alternative hypothesis that \textit{Capotauro} is a pair-instability supernova (PISN) originating from a massive ($\sim250$--$260\,M_\odot$), metal-free star. Using state-of-the-art PISN light curves, spectral energy distributions, and synthetic spectra, we show that a PISN at $z\simeq 15$ can plausibly reproduce the observed brightness, temporal evolution, photometry, and NIRSpec spectrum. We compare this scenario with alternative interpretations, including a local Y0 brown dwarf, and discuss observational tests to discriminate among them. If confirmed, this event would provide a rare window onto Population~III stars, and highlights the importance of transient contamination in ultra-high redshift galaxy samples.
\end{abstract}

\keywords{galaxies: high-redshift, galaxies: evolution, galaxies: formation}

\maketitle

\section{Introduction} \label{sec:intro}
Since the beginning of its operation, the James Webb Space Telescope (JWST) has provided data of unprecedented quality. Their analysis has revealed a population of bright, blue galaxies at very high redshifts ($z>10$), the so-called ``Blue Monsters'' \citep{Naidu22, Haro23, Hsiao23, Wang23, Fujimoto23b, Atek22, Curtis23, Bunker23, Tacchella23, Haro23b, Finkelstein23, Castellano24, Zavala24, Helton24, Robertson24}. These sources are more luminous and common than predicted by virtually all pre-JWST models.

Given that the earliest data were primarily photometric, concerns were raised about the presence of low-redshift contaminants in the samples. A prime example of such a misidentified candidate is CEERS-93316, which initially had a photometric redshift estimate of $z \approx 16$ \citep{Donnan23a} but was later spectroscopically confirmed to be at $z = 4.9$. This discrepancy arose from strong emission lines that mimicked the expected colors of a much more distant object \citep{Haro23b}. However, aside from a few extreme cases like CEERS-93316, the spectroscopic confirmation rate has been remarkably high ($\approx 80\%$, \citealt[][]{Roberts24, Carniani24a, Castellano25}), thus solidifying the tension between theory and observations.

In response, theoretical explanations and adjustments were promptly proposed. These include attenuation-free (AFM, \citealt[][]{Ferrara23, Ziparo23, Fiore23, Ferrara24a, Ferrara24b}), feedback-free (FFB, \citealt{Dekel23,Li24}), and stochastic star formation models \citep{Mason23,Mirocha23,Pallottini23}, as well as density-modulated star formation \citep[][]{Somerville25}, and a top-heavy stellar initial mass function \citep[][but see \citealt{Cueto24}]{Inayoshi22}. Despite these efforts, as observations pushed to increasingly higher redshifts, the discrepancy has persisted and grown, even when compared to the predictions of updated models.

This diverging trend is partly driven by the spectroscopic confirmation of the two most distant known galaxies at $z\approx 14$ \citep[][]{Carniani24a, Naidu25}. These galaxies, with $M_{\rm UV} < -20$, confirm a surprisingly high number density, $\phi_{\rm UV} = 10^{-5.36}\ \rm Mpc^{-3}$, of bright galaxies at such early cosmic epochs. Although some models, like the AFM, remain within $1\sigma$ of the current data, the predicted UV luminosity density, $\rho_{\rm UV}(z)$, declines with redshift more rapidly than observed.

Additionally, if some of the photometric candidates reported by \cite{Perez25, Castellano25, Gandolfi25b} in the redshift range $15 < z < 28$ are confirmed, the inferred $\rho_{\rm UV}(z)$ would exceed AFM predictions at $z=17$ by a factor of $\approx 10$. By $z=25$, matching such a high density of galaxies would become virtually impossible for any current model of galaxy formation.

\begin{figure*}
\centering\includegraphics[width = 1.0\textwidth]{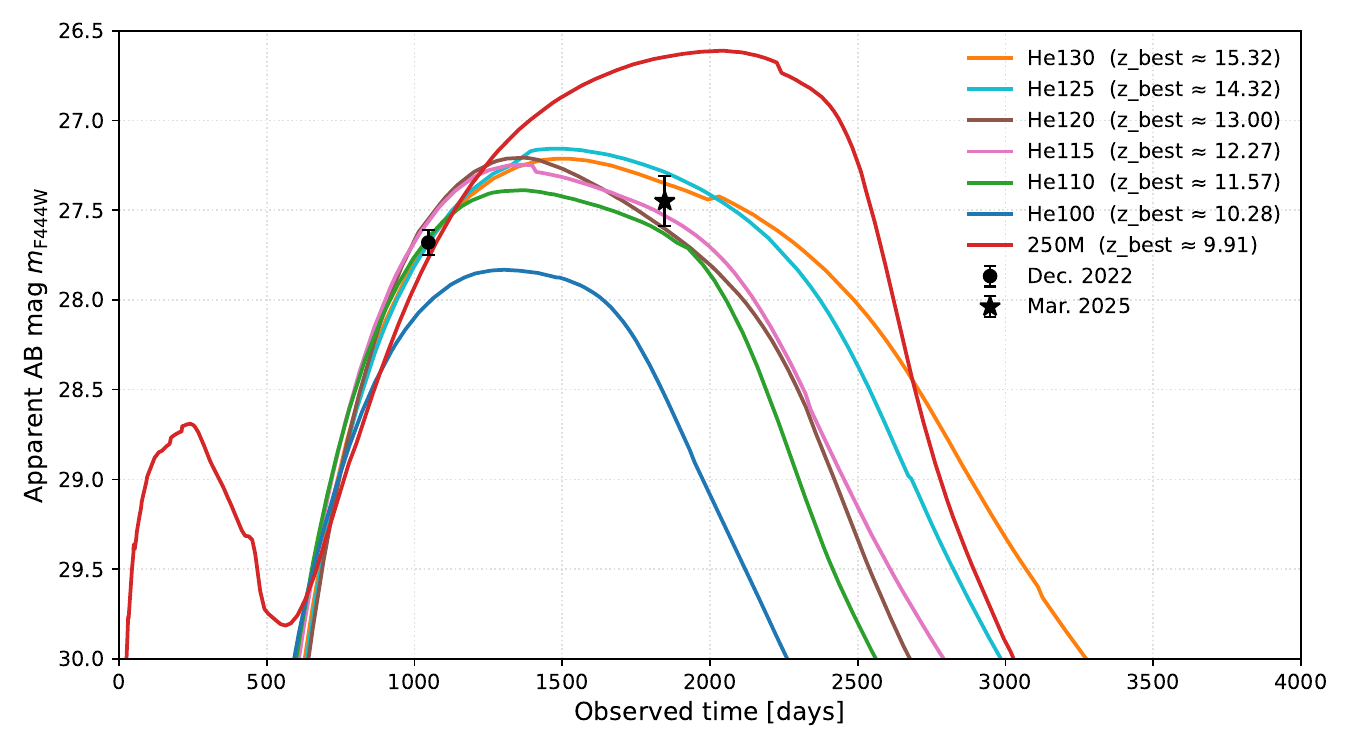}
\caption{Best-fit redshift light curves at $4.44\ \mu$m for different PISN models as a function of the observed time from the explosion. The data points with errors are the NIRCam (circle, 1st epoch) and NIRSpec (star, 2nd epoch) observations \citep{Gandolfi25}.}
\label{fig:light_curves}
\end{figure*}

Should these $z>15$ photometric candidates be confirmed, it \textbf{might} become necessary to at least partially tweak the concordance $\Lambda$CDM model\footnote{Throughout the paper, we assume a flat Universe with the following cosmological parameters: $\Omega_m = 0.3075$, $\Omega_{\Lambda} = 1- \Omega_{\rm M}$, and $\Omega_{b} = 0.0486$, $h=0.6774$, $\sigma_8=0.826$, where $\Omega_{m}$, $\Omega_{\Lambda}$, and $\Omega_{b}$ are the total matter, vacuum, and baryon densities, in units of the critical density; $h$ is the Hubble constant in units of $100,\kms \rm Mpc^{-1}$, and $\sigma_8$ is the late-time fluctuation amplitude parameter \citep{planck:2015}.} by introducing more extreme hypotheses. One of the most fascinating solutions is to postulate that ultra-early galaxies are powered by accreting primordial black holes \citep[][]{Liu23, Matteri25a, Matteri25b}. While this idea is tantalizing and successfully explains the evolution of $\rho_{\rm UV}$ out to $z\approx 30$, its implications still need to be thoroughly explored. Additional suggestions include an early dark energy contribution at $z \approx 3400$ \citep[][]{Shen24}, a tilted primordial power spectrum \citep[][]{Parashari23}, and the presence of non-Gaussianities \citep[][]{Biagetti23}.

The recent discovery of a very bright putative galaxy at $z\approx 30$, dubbed \textit{Capotauro} in \citet{Gandolfi25}, further worsens the problem. \textit{Capotauro} is an F356W-dropout identified in the CEERS survey with an F444W AB magnitude of $27.68$. These properties would place it at $z\approx 30$ with a dazzling $M_{\rm UV} \sim -21.4$ and a number density of $\approx 2\times 10^{-6}\ \rm Mpc^{-3}$.

\textit{Capotauro} has been observed at a second epoch (approximately 800 days after the CEERS NIRCam observations) with NIRSpec as part of the CAPERS survey (P.I. M. Dickinson; \citealt{Donnan25}). Strikingly, the source shows clear signs of variability, having brightened by about 0.23 mag at 4.4 $\mu$m.

Prompted by this unexpected result and considering the aforementioned difficulties of standard galaxy formation scenarios, we explore here the possibility that \textit{Capotauro} is instead a pair-instability supernova (PISN).

The prospect of detecting high-$z$ PISNe, a central aim of this work, is supported by two decades of theoretical research. Pioneering studies by \citet{Scannapieco05} and \citet{Wise05} established PISNe as compelling tracers of primordial star formation and the Pop III initial mass function (IMF), garnering significant subsequent attention \citep{Pan12, Hummel12, Whalen13, Meiksin13, deSouza13, Smidt15, Lazar22}. A distinct line of research has focused explicitly on their detectability, with numerous works exploring predicted rates and observational signatures \citep{Hartwig18, Moriya19, Regos20, Moriya21, Yan23, Gabrielli24, Jeon26}. While these models vary considerably in their predicted event rates, a consensus emerges that PISNe occurring at Cosmic Dawn and during the Epoch of Reionization should be detectable in appreciable numbers within the \textit{JWST} observational lifetime.

\begin{figure*}
\centering\includegraphics[width = 1.0\textwidth]{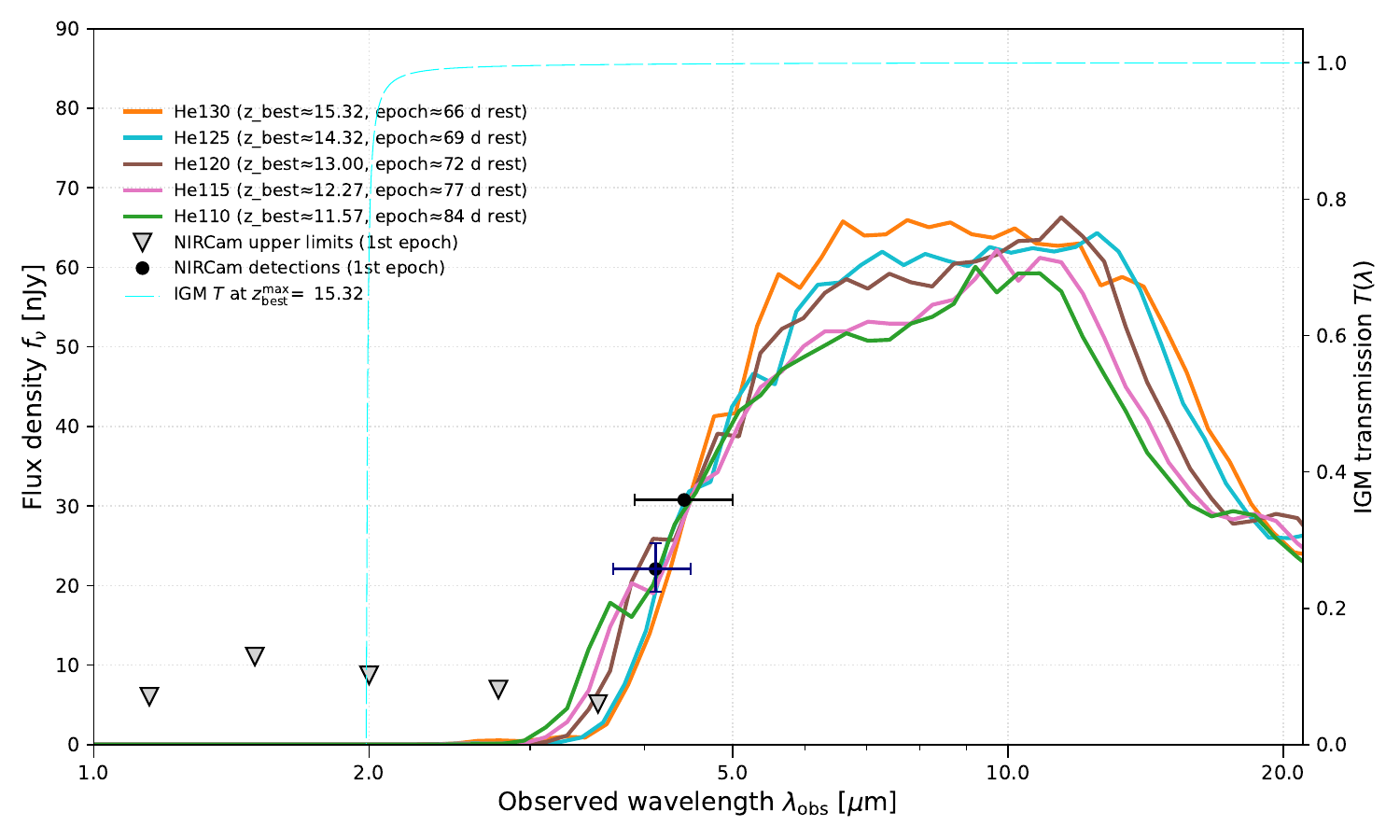}
\caption{Spectral energy distributions of the five fiducial models fitting the light curve constraints as a function of the observed wavelength at the 1st observation epoch (Dec. 2022). The curves are color-coded as shown in the label. Also shown are the two measured NIRCam data points (black circles), and upper limits (gray triangles) \citep[][]{Gandolfi25}. For illustration, we show the transmissivity (cyan dashed line) due to the Ly$\alpha$ damping wing at the maximum redshift of the models ($z=15.32$); its effects on the SEDs are completely negligible.}
\label{fig:SED}
\end{figure*}

\section{Data} \label{sec:data}
The discovery of \textit{Capotauro} has been reported in \citet[][]{Gandolfi25}. We refer to that paper for a detailed description of the observations and of the data reduction. Here we simply summarize some key aspects that are crucial for this work.

The source has been observed at two different epochs. In the first epoch (December 2022) it has been imaged as part of the CEERS Program (P.I. S. Finkelstein; \citealt{Finkelstein25}) in 8 NIRCam bands; these data were complemented with 6 HST bands. The source is a F356W-dropout and was detected in the two filters F444W (flux density $f_\nu = 30.7 \pm 2.4\ \rm nJy$) and F410M ($f_\nu = 22.1 \pm 4.5\ \rm nJy$). On this basis, while exploring different possibilites including brown dwarfs, low-$z$ dusty galaxies, free-floating planets or Little Red Dots, among the extragalactic options, \citet[][]{Gandolfi25} seem to favor the interpretation \textit{Capotauro} as a $z\approx 30$ galaxy candidate.

After $\sim 800$ days (March 2025) a JWST/NIRSpec PRISM/CLEAR spectrum of \textit{Capotauro} was acquired as part of the CAPERS Program (P.I. M. Dickinson; see, e.g., \citealt{Donnan25}). We re-processed the spectroscopic data by using both the STScI Calibration Pipeline version 1.20 and NIRSpec GTO pipeline \citep{DEugenio25, Scholtz25}, adopting the latest calibration reference files (Calibration Reference Data System Context 1464). The calibrated spectrum has been extracted using a PSF-based optimal extraction aperture to maximize the signal-to-noise ratio. The spectra obtained from the two pipelines are consistent with each other and in agreement with that reported by \citet{Gandolfi25}. We therefore adopt the spectrum produced by the NIRSpec GTO pipeline as our fiducial dataset. Finally, we derived the synthetic NIRCam photometry at $4.4\mu$m by convolving the spectrum with the nominal F444W filter throughput, and propagating the uncertainties from each spectral channel.

The 2nd epoch measured flux density is $f_\nu = 37.8 \pm 4.7\ \rm nJy$, corresponding to an apparent magnitude $m_{\rm AB} = 27.45 \pm 0.14$. Thus, the source between the two epochs has brightened by $\simeq 20\%$.

CEERS re-observed the source with MIRI in 4 bands in March 2024, i.e. about one year before the 2nd epoch. Only upper limits have been obtained. In the deepest filter, F777W, they got a 3$\sigma$ flux upper limit $< 191$ nJy. Hence, these data are not particularly constraining for our purpose, but they can be used as model sanity checks.

Flux calibration, and thus path-loss correction, of NIRSpec data is accurate to the 5--10\% level for point-like sources, depending on the target position within the micro shutter, with reduced accuracy ($\sim20\%$) for extended emission \citep{Scholtz25}. In this case, \textit{Capotauro} appears sufficiently compact to be treated as point-like. Under this assumption, the discrepancy between the NIRCam and NIRSpec measurements suggests a difference at the $>2\sigma$ level, although residual calibration uncertainties cannot be excluded.

If confirmed, the observed variability can be explained by a transient phenomenon. Concerning the former, the brightness of the source excludes standard core-collapse supernovae. However, a class of very luminous PISNe is expected to originate from metal-free, massive ($140-260 M_\odot$) stars \citep[][]{Heger02, Kasen11, Whalen13, Chen14} that should become more common at high redshift. In the following we explore whether a PISN can explain both the time variability and the spectrum of \textit{Capotauro}.

\begin{figure*}
\centering\includegraphics[width = 1.0\textwidth]{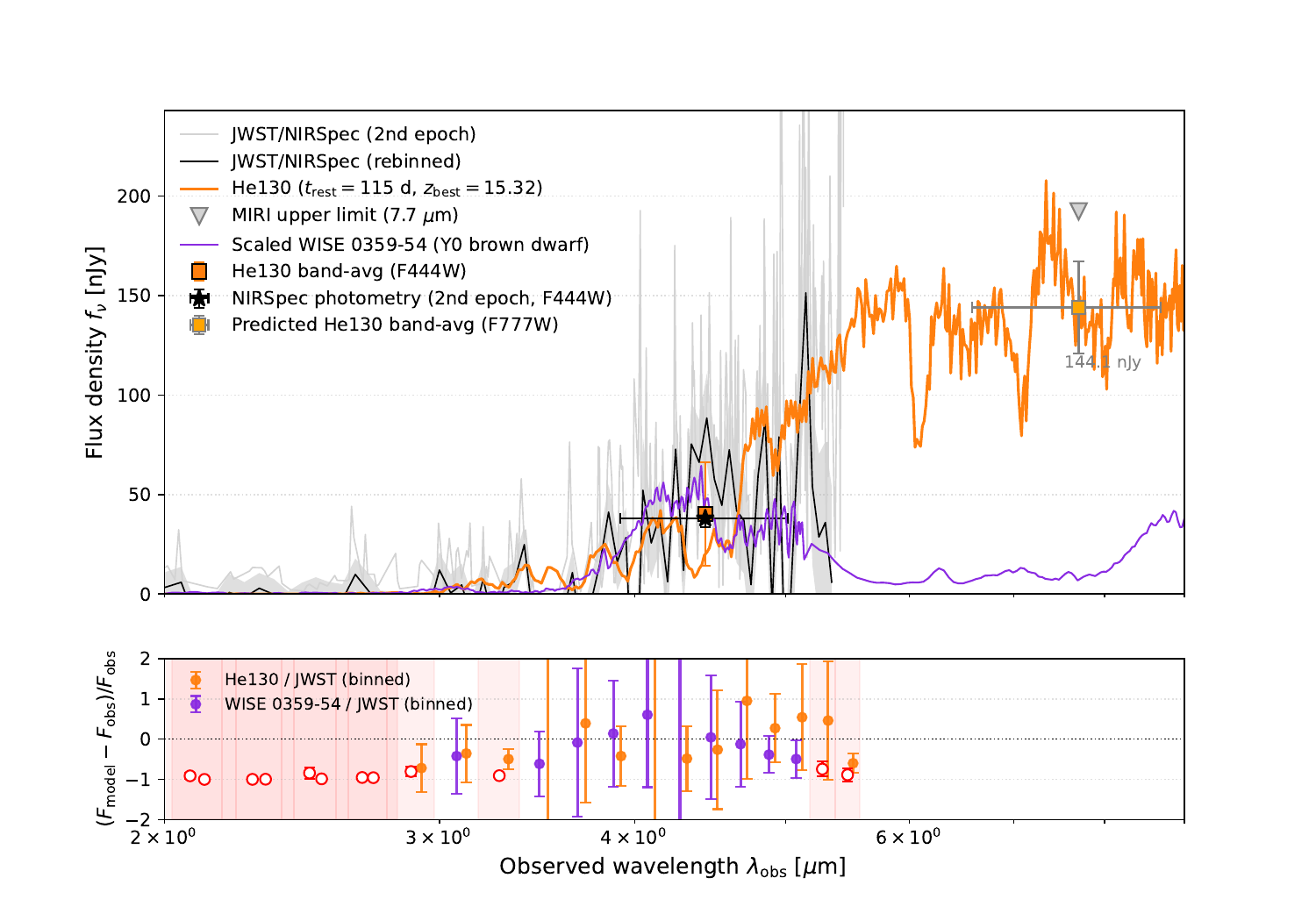}
\caption{\textit{Top panel}: Comparison between \textit{Capotauro} spectrum (gray line; the black line shows the rebinned spectrum) and the He130 PISN model (orange) shown at its best-redshift ($z=15.32$), and 115 day after the explosion, i.e. at the 2nd epoch. The orange squares are the predicted photometric flux densities in the F444W (observed value shown as black star), and F777W filters. Also shown is the CEERS MIRI F777W 3$\sigma$ upper limit \citep[][]{Gandolfi25}. The spectrum of a prototypical cool Y0 brown dwarf (WISE 0359-54, \citealt{Beiler23, Beiler24}), scaled to match the F444W photometric point, is shown as a purple line. \textit{Bottom}: Fractional binned residuals of the PISN (orange filled points) and brown dwarf spectra (purple filled) after subtraction of the JWST spectrum. Open red points/light red bands indicate a poor fit, i.e. bins where the mean is $>3\sigma$.}
\label{fig:spectra}
\end{figure*}

\section{PISN light curves} \label{sec:pisn}
As a first step, we check whether theoretical PISN light curves can explain the observed brightness and time-variability of \textit{Capotauro}. To this aim we use the data in the Garching Core-Collapse Supernova Archive\footnote{\url{https://wwwmpa.mpa-garching.mpg.de/ccsnarchive/}} which contains a wavelength-dependent PISN light curve collection. These light curves (S. Kozyreva, private comm.) have been computed using the hydrodynamics radiative-transfer code STELLA \citep[][]{Blinnikov98}. We note that the observed light curves are time-dilated by a factor $(1+z)$ if the explosion occurs at redshift $z$.

The study is based on stellar models, presented in \citet[][]{Heger02}, describing the explosive nucleosynthesis of helium cores in the mass range $M_{\rm He} = 65-130\ M_\odot$, corresponding to zero-age main-sequence star masses $M_{\rm ZAMS} = (24/13) M_{\rm He} + 20$ (all masses in solar units). Thus, models span the entire PISN range of $130-260\ M_\odot$, and are classified according to $M_{\rm He}$ as HeX, with X in the range $65-130$. The $\rm ^{56}Ni$ mass powering the luminosity of the PISN increases from $0.03 M_\odot \to 40 M_\odot$ in the above stellar mass range. The peak bolometric luminosity of the most massive model, He$130$, whose explosion energy is $8.67\times 10^{52}\ \rm erg\ s^{-1}$, exceeds $10^{44}\ \rm erg\ s^{-1}$. The peak is reached $\approx 100$ days after explosion by essentially all models, but the peak luminosity decreases for smaller values of the core mass.

In addition to the fiducial HeX models, in which stars have lost their hydrogen envelopes to mass loss or binary mass exchange just prior to exploding, we have also experimented with ``hydrogenic'' models in which the star still retains its H-envelope. These are presented for metal-free and $Z=10^{-4} Z_\odot$ in \citet[][]{Kasen11}. The primary difference between the zero-metallicity pre-PISN stars and those with a metallicity of $Z=10^{-4} Z_\odot$ solar is that the former end their lives as compact blue supergiants, whereas the latter evolve into red supergiants.

Finally, we also considered two other progenitor models from \citet[][]{Kozyreva14}, i.e. a $150\ M_\odot$ (named 150M) red supergiant, and a $250\ M_\odot$ (250M) yellow supergiant at a metallicity of $Z = 0.001$. These models also follow the initial break-out of the shock from the envelope; they are considered to be fair analogues of candidate low-redshift PISNe.

Armed with this sample of light curves, we compared them with the \textit{Capotauro} photometric data at the two different epochs (Fig.~\ref{fig:light_curves}). First we note that the most luminous bare helium core models (He110--He130) are all capable to provide acceptable fits to the observations. On the contrary, less massive progenitors and hydrogenic models fail to match the observed luminosity and/or time-evolution of the variability.

To elucidate this failure we have shown the He100 and the 250M models as an example. For this reason in the following we will concentrate on the five most promising \textit{fiducial} models, He110, He115, He120, He125, and He130. These are also shown in Fig.~\ref{fig:light_curves}.

When comparing simulated light curves with the data we have two degrees of freedom: (a) the redshift, $z$ of the PISN, and (b) the time from the explosion corresponding to 1st-epoch observations of the source. The redshift affects both the observed luminosity and the light curve time span. As the source is brightening between the first and second epoch, the time from explosion is set by the first time at which the PISN reaches the observed F444W magnitude ($m_{\rm F444W}= 27.68$). We then determined the actual redshift of the PISN via a best-fitting procedure.

Light curves raise very rapid and similarly for the five fiducial models; however, they differ significantly in the dimming phase. For this reason, independently of the model, the first observed epoch coincides with $\approx 1000$ days after explosion. All models are almost equally successful in reproducing the 1st and 2nd epoch data. The main difference is that more massive PISNe are located at progressively higher redshifts, ranging from $z=11.57$ (He110) to $z=15.32$ (He130). Note that this range is well below the redshift $z\approx 32$, necessary to interpret \textit{Capotauro} as a ultra-early galaxy.

The light curve analysis alone is not sufficient to break the degeneracy among the five fiducial models. To this aim we need to analyse their Spectral Energy Distribution at the relevant observation epoch.

\section{Spectral Energy Distributions} \label{sec:SED}
The SEDs of the five fiducial models, each shown at their best-fit redshift, are presented in Fig.~\ref{fig:SED} as a function of the observed wavelength for the first epoch observation (Dec. 2022). These are compared with the two observed photometric points and the upper limits in the other JWST bands.

The PISN SEDs drop shortwards of $3-3.5\ \mu$m, corresponding to a r.f. wavelength of $\approx 2000-2500\ \rm \AA$ depending on the model. This drop is caused by the strong absorption of UV photons occurring in the ejecta. Such short-wavelength cut explains why the redshift of the putative PISN is lower than that ($z\approx 32$, \citealt{Gandolfi25}) inferred for the spectrum used for \textit{Capotauro}.

In Fig.~\ref{fig:SED} we also plot the Ly$\alpha$ damping wing due to intergalactic HI scattering at the maximum redshift ($z=15.32$) inferred for the models. While the damping wing could be potentially important as the cosmic gas is neutral at that epoch, the PISN SED is observed sufficiently away from $1215 (1+z)\ \rm \AA$ that it is not affected by such process.

All fiducial models match quite well the F410M and F444W data points. This is not surprising as they have been singled-out as the best light curves at $4.44\ \mu$m. However, only the two most massive models (He125 and He130) satisfy the 1$\sigma$ upper bound ($< 3.2\ \rm nJy$) set by the F356W filter. The other three models are excluded also by the $3\sigma$ upper limit. Other filters, including the F770W, do not significantly constrain the models. The two successful models are virtually indistinguishable in the entire NIRCam sensitivity range, and they differ by $< 10\%$ even at $7.7\ \mu$m. In the following, then, we will concentrate on the analysis of the He130 model.

\section{Spectra} \label{sec:spectra}
As a final step, in the top panel of Fig.~\ref{fig:spectra} we compare the predicted spectrum of the He130 PISN model taken from \citet{Kasen11} at $z = 15.32$ (orange line) with the spectrum observed by NIRSpec at the second epoch\footnote{Approximately 115 days after explosion in the PISN rest frame.} (March 2025; gray). The rebinned spectrum (black) is also shown to facilitate the comparison.

As already noted in the discussion of the light curves in Fig.~\ref{fig:light_curves}, the He130 model provides an excellent fit to the NIRSpec F444W photometry. The PISN spectrum exhibits an overall rise with wavelength up to $\approx 6\ \mu$m, flattening at longer wavelengths. Superimposed on this continuum trend are several absorption features produced by freshly synthesized intermediate-mass elements. These include iron-group line blanketing at shorter wavelengths and the prominent Mg I and Mg II absorption features visible at $6$--$8\ \mu$m. When averaged over the F777W filter, the predicted flux is $144.1\ \rm nJy$. This is still well below the CEERS upper limit\footnote{This data point has been obtained 1 yr before the 2nd epoch} ($< 191$ nJy), but it could be potentially detectable with MIRI.

The bottom panel of Fig.~\ref{fig:spectra} shows the fractional (binned) residuals obtained by subtracting the observed spectrum from the predicted one. The residuals are generally small ($\lesssim 50\%$) and within the (large) error bars, indicating that the He130 PISN spectrum provides a good match to the \textit{Capotauro} spectrum. The largest discrepancy is found at $4.1\ \mu$m, where uncertainties related to iron-group line blanketing are significant. These features are, in fact, highly sensitive to the exact epoch of observation relative to the explosion.

Although the observed variability, if confirmed, would strongly favour a transient origin of \textit{Capotauro}, such as a PISN, we nevertheless consider also the alternative possibility that the source is a local brown dwarf (BD).

The shape and amplitude of the observed spectrum already limits this interpretation to a specific class of BDs, i.e. Y dwarfs. These are the coolest products of star formation, with effective temperatures $< 600$ K \citep[][]{Kirkpatrick21, Beiler24} with spectra peaking at 5 $\mu$m. The prototype of this class is WISE 0359-54 \citep{Beiler23}, a cool (effective temperature $\sim 450$ K) Y0 BD located at a distance of 13.57 pc. In Fig.~\ref{fig:spectra} we report its spectrum (purple line, \citealt{Beiler23}) with amplitude rescaled to match the F444W flux (38.02 nJy). This is equivalent to state that the putative BD identified with \textit{Capotauro} is located at an actual distance of 785 pc. As it can be seen from the residuals in the bottom panel, the spectrum of WISE 0359-54 provides an equally good fit to the data as the PISN. The main difference is that, contrary to the He130 model, the BD spectrum drops beyond 5 $\mu$m. Hence, a deeper MIRI photometry and/or NIRSpec spectrum are essential to discriminate between the two hypotheses. We restate that the BD interpretation can however not account for the variability of the source between the two currently available epochs, that should be therefore explained differently.

Finally, we note that the upper limit on the proper motion of \textit{Capotauro} from the 2-epoch observations is $\mu \le 0.137''\ \rm yr^{-1}$. If the BD is located at 785 pc, and therefore approximately shares the tangential velocity of the Milky Way halo ($\approx 200\ \kms$), we would expect a proper motion of $0.053''\ \rm yr^{-1}$, i.e. still compatible with existing observations. A second epoch with NIRCam F444W should be able to confirm or rule out such a proper motion.

\section{Discussion} \label{sec:discussion}
We have shown that a massive ($\sim 250$--$260\ M_\odot$) pair-instability supernova exploding at $z \simeq 15$ can plausibly account for the observed brightness, variability, spectral energy distribution, and spectrum of \textit{Capotauro}. This interpretation can be tested and further constrained through several complementary observational strategies.

\subsection{Testing the PISN hypothesis}
First, confirming the source magnitude variability and assessing its consistency with the light curves of the best-fitting PISN models (He130/He125) would be essential; to minimize potential cross-calibration uncertainties, such observations should ideally be obtained with NIRCam. Second, because PISN spectra are predicted to continue rising beyond the NIRCam wavelength range, deep MIRI photometry could provide an important test of this spectral behavior. Finally, high-resolution JWST spectroscopy would allow a more robust identification of the characteristic absorption features expected in PISN models.

Taken together, these observations would not only provide strong support for the interpretation of \textit{Capotauro} as an extremely high-redshift supernova, but could also offer a rare observational window into the final stages of Population III stellar evolution. More broadly, such a detection would have important implications for our understanding of star formation and galaxy assembly at ultra--high redshift, a regime that remains challenging to reconcile with current cosmological models. Future work, applying the same methodology introduced here, will clarify whether other potentially ultra-high redshift sources \citep[e.g.][]{Perez25,Castellano25,Kokorev25} could instead arise from PISN transients.

Importantly, these studies would naturally complement more dedicated campaigns, such as the two-epoch JADES \citep[][]{Eisenstein23} + CONGRESS (PI: Egami; JWST Proposal \#6541) transient survey \citep[][]{deCoursey25}. In addition to JADES, many deep JWST imaging programs (PRIMER, GLASS-JWST, UNCOVER, NGDEEP, CEERS, COSMOS-Web) -- though not explicitly designed only for transients -- provide the necessary data depth and repeat observations where PISNe could be found serendipitously.

\subsection{Expected PISN rate in CEERS}
As a sanity check of the PISN interpretation we estimate the expected rate of PISNe in the CEERS survey volume in which \textit{Capotauro} has been discovered. CEERS covers $\sim 90\ \rm arcmin^2$ of the Extended Groth Strip, corresponding to a volume $V\sim 10^6\ \rm Mpc^3$ at $10 < z < 20$.

Let us assume a $(0.1 -1000) M_\odot$ Larson IMF $\propto m^{-2.35} \exp(-m_{\rm ch}/m)$ with a characteristic mass $m_{\rm ch}=10\ M_\odot$ \citep[][]{Pagnini23}. Then the number of PISNe per unit stellar mass formed is $\eta_{\rm PISN} = 5.69 \times 10^{-4} M_\odot^{-1}$. We further approximate the Pop III comoving cosmic star formation rate density in $10 < z < 20$ as $\dot \rho_{\rm *,III}(z)=10^{-4}[(1+z)/12]^{-11.3}\ \rm M_\odot\ yr^{-1}\ Mpc^{-3}$ from fitting the results given in \citet[][]{Pallottini14, Venditti23}. The PISN rate is then ${\cal R}(z) = \eta_{\rm PISN} \dot \rho_{\rm *,III}(z)$. If the PISN remains detectable in the restframe for a time $T\approx 100$ days, the expected number of PISNe visible at a random epoch in CEERS is
\begin{equation}
    N = \int_{10}^{20} {\cal R}(z) T \frac{dV}{dz}dz = 6.2 \times 10^{-3}.
\end{equation}
So CEERS would contain a detectable $10<z<20$ PISN about 0.6\% of the time under these assumptions. Although small and rather uncertain, this value does not exclude that \textit{Capotauro} could indeed be a PISN. Interestingly, recent simulation-based PISN rate predictions by \citet[][]{Jeon26} are even more optimistic (see their Fig. 2), predicting almost exactly 1 PISN/yr in the CEERS volume at $z\approx 20$.

\begin{figure}
\centering\includegraphics[width = 0.45\textwidth]{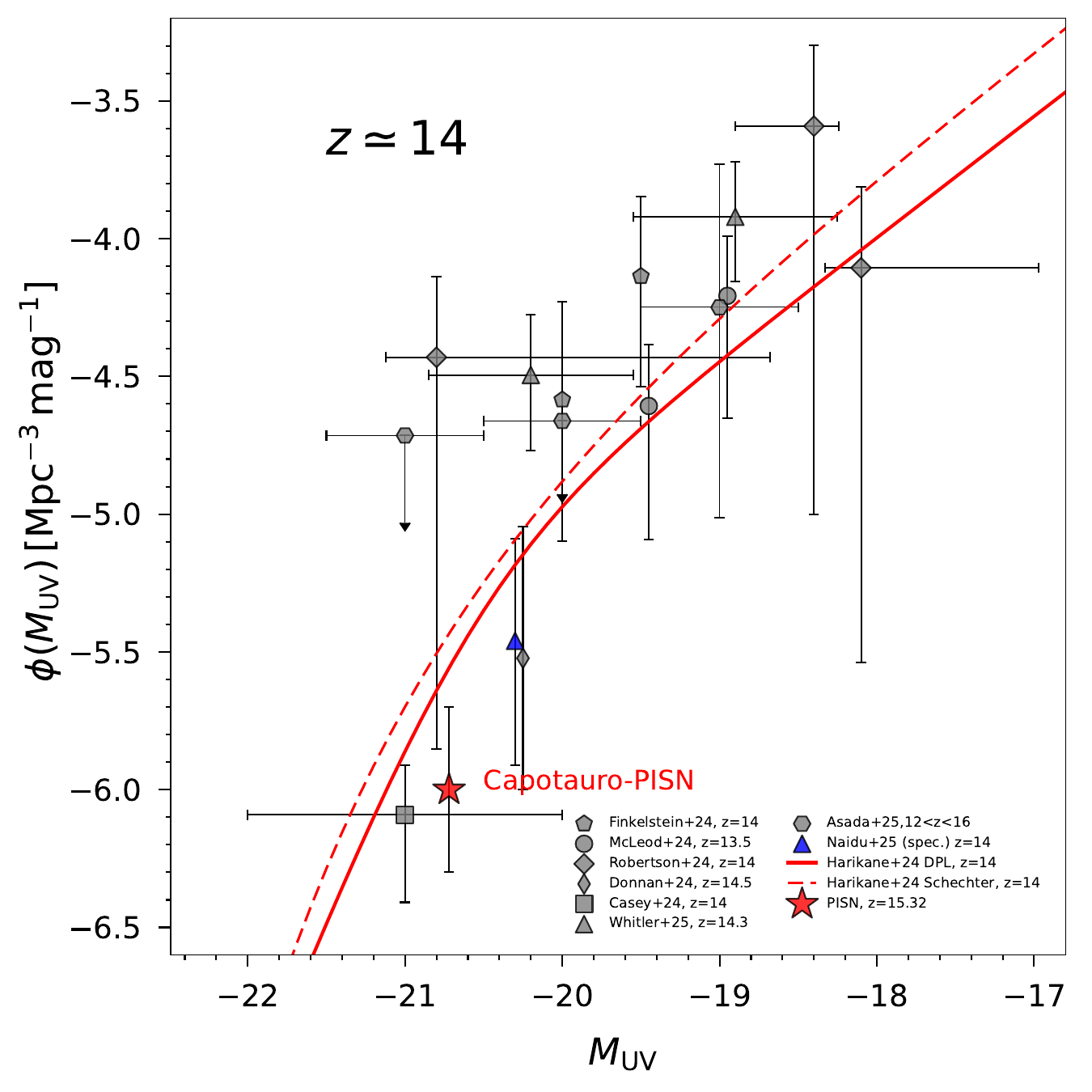}
\caption{UV luminosity function at $z\simeq 14$ showing the putative location of \textit{Capotauro} interpreted as a PISN at $z\simeq 15$. Data points are from \citet[][]{Finkelstein24, McLeod24, Robertson24, Donnan24, Casey_2024, Whitler25, Asada25, Naidu25}; the lines show the fit provided by \citet[][]{Harikane25} adopting a double-power law (solid) or a Schechter (dashed) function. The error on the PISN data point assumes a factor 2 uncertainty in the comoving volume.}
\label{fig:UVLF}
\end{figure}

We can estimate the probability to find a PISN in the survey volume in a different way. It has been shown that, due to inefficient fragmentation, in small halos at cosmic dawn there tends to be one (or at most a few) Pop III stars per halo \citep{Ciardi05, Trenti09, Haemmerle20, Klessen23}. Very crudely, then, the number of PISNe can be deduced from the number density, $n$, of halos at $z\approx 15$, i.e. $N = (T/t_{\rm ff}) n V$, where the free-fall time is related to the Hubble time via $t_{\rm ff}=0.06 t_H(z=15)$ \citep[][]{Ferrara24a}. By setting $N\ge 1$, we find $n> 60\ \rm Mpc^{-3}$, which is satisfied by halos with $M < 10^{6.4}\ M_\odot$, or virial temperature $T_{\rm vir} \simlt 3000$ K. These properties match the expected ones for halos hosting the first PopIII stars \citep[][]{Tegmark97, Abel00, Bromm01, Salvadori09}. Although very rough, this estimate shows that it is not unplausible that with \textit{Capotauro} we have observed one of the first cosmic episodes of PopIII star formation.

We finally note that, with its near-UV ($\lambda = 2000-3800$ \AA) magnitude $M_{\rm UV}=-20.7$, and a number density $n \simeq 10^{-6}\ \rm Mpc^{-3}$, the \textit{Capotauro}-PISN lies very close\footnote{We reiterate that if \textit{Capotauro}-PISN would be interpreted instead as a galaxy, its Lyman break would put it at $z\approx 30$. Hence, it would \textit{not} represent a contaminant of the $z=14$ UV LF.} to the currently available UV galaxy luminosity function (LF) at $z\approx 14$, as shown in Fig.~\ref{fig:UVLF}. This is essentially the most distant redshift for which a relatively solid -- yet almost entirely photometric -- LF is available. We conclude that PISNe might be as common as the brightest galaxies at $z>10$, thus opening exciting perspectives for their detection in dedicated transient surveys.

\section{Summary} \label{sec:summary}
Galaxy formation theory within the standard cosmological framework is facing increasing difficulty in explaining the presence of bright sources at ultra-early epochs ($z>15$). Aside from invoking drastic modifications to the $\Lambda$CDM model or contamination by lower-redshift interlopers, proposed solutions include non-stellar sources, such as accreting (primordial) black holes \citep[][]{Liu23, Matteri25a, Matteri25b}, or transient phenomena.

In this work, we have focused on \textit{Capotauro}, a very red source originally interpreted as a galaxy at $z\approx 30$, which exhibits clear evidence of variability over an $\sim800$-day baseline. We have shown that its variability, spectral energy distribution, and NIRSpec spectrum can be consistently reproduced if the source is instead identified as a metal-free pair-instability supernova originating from a $250$--$260\ M_\odot$ progenitor at $z\simeq 15$.

We find that an equally good spectral fit can be obtained by modeling \textit{Capotauro} as a cool ($\sim450$ K) Y0 brown dwarf located at a distance of $\sim785$ pc; however, in this scenario the observed variability remains unexplained.

Further progress requires confirming the variability and proper motion and testing its consistency with the light curves of the best-fitting PISN models (He125/He130). Deep MIRI photometry and high-resolution JWST spectroscopy will be crucial to discriminate between PISN and brown dwarf interpretations.

If confirmed, the PISN scenario would not only identify \textit{Capotauro} as the most distant supernova observed to date, but would also provide a rare observational window onto the final stages of Pop~III stellar evolution. Crucially, even a single detection of PISN can be crucial to our understanding of the mass distribution of the first stars \citep[][]{Koutsouridou24}.

\section*{Acknowledgments}
We thank A. Kozyreva, D. Kasen, A. Heger, S. Beiler for providing key data, G. Rodighiero, I. Labb\'e, G. Gandolfi, S. Salvadori for useful discussions and sharing data. This work is supported by the ERC Advanced Grant INTERSTELLAR H2020/740120, and in part by grant NSF PHY-2309135 to the Kavli Institute for Theoretical Physics. MS acknowledges the hospitality of the Scuola Normale Superiore in Pisa (Italy) where some of this work took place and partial support through NASA grant 80NSSC21K1294. SC acknowledges support from the European Union (ERC, WINGS,101040227). This work made use of data from the MPA Archive \url{https://wwwmpa.mpa-garching.mpg.de/ccsnarchive/}.


\bibliographystyle{aasjournal}
\bibliography{paper}

\end{document}